\documentclass[aps, reprint, floatfix]{revtex4-2}

\usepackage{graphicx}  
\usepackage{dcolumn}  
\usepackage{bm}  
\usepackage{hyperref}  
\usepackage{amssymb}  
\usepackage{amsmath}  
\usepackage{braket}
\usepackage{times}
\usepackage{units}

\begin{document}

\title{Interferometry of Atomic Matter Waves in the Cold Atom Lab\\ onboard the International Space Station} 

\author {Jason R. Williams,$^{1\ast}$ Charles A. Sackett,$^{2\ast}$ Holger Ahlers,$^{3}$ David C. Aveline,$^{1}$ \\Patrick Boegel,$^{4}$ Sofia Botsi,$^{1}$ Eric Charron,$^{5}$ Ethan R. Elliott,$^{1}$ Naceur Gaaloul,$^{6}$\\ Enno Giese,$^{7,8}$ Waldemar Herr,$^{3}$ James R. Kellogg,$^{1}$ James M. Kohel,$^{1}$ Norman E. Lay,$^{1}$\\ Matthias Meister,$^{10}$ Gabriel M\"{u}ller,$^{6}$ Holger M\"{u}ller,$^{9}$ Kamal Oudrhiri,$^{1}$ Leah Phillips,$^{1}$\\ Annie Pichery,$^{5,6}$ Ernst M. Rasel,$^{6}$ Albert Roura,$^{10}$ Matteo Sbroscia,$^{1}$\\ Wolfgang P. Schleich,$^{4,11,12,13}$ Christian Schneider,$^{1}$ Christian Schubert,$^{3}$ \\Bejoy Sen,$^{2}$
Robert J. Thompson,$^{1}$ Nicholas P. Bigelow $^{14\ast}$\\
\normalsize{$^{1}$Jet Propulsion Laboratory, California Institute of Technology, Pasadena, CA 91109, USA}\\
\normalsize{$^{2}$Department of Physics, University of Virginia, Charlottesville, Virginia 22904, USA}\\
\normalsize{$^{3}$German Aerospace Center (DLR), Institute of Satellite Geodesy and Inertial Sensing, 30167 Hannover, Germany}\\
\normalsize{$^{4}$Institut f\"{u}r Quantenphysik and Center for Integrated Quantum Science and Technology (IQST),}\\
\normalsize{Ulm University, Ulm, Germany}\\
\normalsize{$^{5}$Université Paris-Saclay, CNRS, Institut des Sciences Moléculaires d’Orsay, F-91405 Orsay, France}\\
\normalsize{$^{6}$Leibniz University Hannover, Institute of Quantum Optics, QUEST-Leibniz Research School, Hanover, Germany}\\
\normalsize{$^{7}$Technische Universit\"at Darmstadt, Fachbereich Physik, Institut f\"{u}r Angewandte Physik, Darmstadt, Germany}\\ 
\normalsize{$^{8}$Institut fur Quantenoptik, Leibniz Universit\"{a}t Hannover, Hannover, Germany}\\
\normalsize{$^{9}$Department of Physics, University of California, Berkeley, CA, USA}\\
\normalsize{$^{10}$German Aerospace Center (DLR),}
\normalsize {Institute of Quantum Technologies,}
\normalsize {89081 Ulm, Germany}\\
\normalsize{$^{11}$Hagler Institute for Advanced Study, Texas A\&M University, College Station, TX, USA}\\
\normalsize{$^{12}$Texas A\&M AgriLife Research, Texas A\&M University, College Station, TX, USA}\\
\normalsize{$^{13}$Institute for Quantum Science and Engineering (IQSE), Department of Physics and Astronomy,}\\
\normalsize{Texas A\&M University, College Station, TX, USA}\\
\normalsize{$^{14}$Department of Physics and Astronomy, Institute of Optics, Center for Coherence and Quantum Optics}\\
\normalsize {University of Rochester, Rochester, New York 14627, USA}\\
\normalsize{$^\ast$Corresponding Authors: jrwillia@jpl.nasa.gov, cas8m@virginia.edu, nbigelow@ur.rochester.edu}
}

\begin{abstract}
Ultracold atomic gases hold unique promise for space science by capitalizing on quantum advantages and extended freefall, afforded in a microgravity environment, to enable next-generation precision sensors. Atom interferometers are a class of quantum sensors which can use freely falling gases of atoms cooled to sub-photon-recoil temperatures to provide unprecedented sensitivities to accelerations, rotations, and gravitational forces, and are currently being developed for space-based applications in gravitational, earth, and planetary sciences, as well as to search for subtle forces that could signify physics beyond General Relativity and the Standard Model. NASA’s Cold Atom Lab (CAL) operates onboard the International Space Station as a multi-user facility for studies of ultra\-cold atoms and to mature quantum technologies, including atom interferometry, in persistent micro\-gravity. In this paper, we report on path-finding experiments utilizing ultra\-cold $^{87}$Rb atoms in the CAL atom interferometer, which was enabled by an on-orbit upgrade of the CAL science module: A three-pulse Mach-Zehnder interferometer was studied to understand limitations from the influence of ISS vibrations. Additionally, Ramsey shear-wave interferometry was used to manifest interference patterns in a single run that were observable for over 150~ms free-expansion time. Finally, the CAL atom interferometer was used to remotely measure the photon recoil from the atom interferometer laser as a demonstration of the first quantum sensor using matter-wave interferometry in space.
\end{abstract}

\maketitle

\section{Introduction}\label{Introduction}
NASA's Cold Atom Lab (CAL) was launched to the International Space Station (ISS) in 2018 as a multi-user facility for fundamental physics investigations with ultracold atoms enabled by a persistent microgravity environment. The flight instrument incorporates a toolbox of capabilities for flight Principal Investigators (PIs) to pursue space-enabled studies with interacting quantum gases \cite{CALug}. Initial commissioning efforts for the flight instrument used laser and evaporative cooling of rubidium gases to produce Bose-Einstein Condensates (BECs) in orbit and characterized these quantum gases after freefall times exceeding 1 second \cite{Aveline2020}. Subsequent science demonstrations by the CAL investigators included enhanced techniques for cooling, manipulation, and control for quantum gases such as decompression cooling via adiabatic relaxation \cite{Pollard2020}, shortcut to adiabaticity and delta-kick collimation techniques for preparing atoms with effective temperatures as low as 52 picokelvin \cite{Gaaloul2022}, as well as the development of RF-dressed potentials and subsequent study of ultracold rubidium gases in bubble-shaped geometries \cite{Carollo2022}. 

Accommodation on the ISS not only provides the unique infrastructure, support, and effective real-time download of science data needed for daily operations of CAL, but also allows for on-orbit repair and upgrades by the resident astronaut crew. Two key upgrades have significantly expanded the scientific capabilities of the flight instrument: (i) A replacement science module (SM-3) was launched to ISS onboard the commercial resupply services mission SpaceX-19 and was installed into CAL by astronauts Christina Koch and Jessica Meir. SM-3 was equipped with a new optical beam path to enable experiments with atom interferometry. (ii) An upgraded microwave source was brought to the ISS onboard SpaceX-22 and was installed into CAL by astronaut Megan McArthur. The enhanced microwave source enabled species-specific evaporative cooling of $^{87}$Rb gases to ultracold temperatures to provide a reservoir for sympathetic cooling of bosonic potassium gases. Recent commissioning efforts of these hardware upgrades have produced ultracold gases of either $^{41}$K or $^{39}$K, coexisting and interacting BECs of $^{41}$K and $^{87}$Rb, and simultaneously interrogated $^{41}$K and $^{87}$Rb atom interferometers using the single-laser atom interferometer beam \cite{Elliott2023}. This article details the design of the CAL atom interferometer upgrade and the first PI-led experiments utilizing matter-wave interferometry with ultracold $^{87}$Rb gases in orbit.

Precision metrology based on light-pulse atom interferometry is a quintessential utilization of the wave-like nature of matter that is at the heart of quantum mechanics. Here, pulses of laser light generate a superposition of motional states of atoms whose momenta differ by discrete units of the photon recoil. Subsequent laser pulses then induce the components of the atomic wave functions to recombine and constructively or destructively interfere (see Methods \ref{Sect:MethodsBraggAI} for more details). The final result is an interference pattern that provides a phase-sensitive readout of effects such as accelerations, rotations, gravity, and subtle forces that could signify new physics acting on matter. Due to the quantum nature of ultracold atomic gases and the phase sensitivity of matter waves in extended free fall, atom interferometry can enable unprecedented applications for both applied and fundamental physics \cite{Bongs2019}. Such applications include gravimetry \cite{Kasevich1992,Wu2019,Freier2016,Altin_2013,7477832} and gravity gradiometry \cite{PhysRevLett.81.971,PhysRevA.65.033608,Yu2006}, investigating gravity at microscopic scales \cite{PhysRevLett.97.060402}, tests of the universality of free fall (UFF) with quantum matter \cite{Zho15,PhysRevLett.125.191101,PhysRevLett.112.203002} and measurement of relativistic effects with delocalized quantum superpositions \cite{Loriani2019,Roura2020,Ufrecht2020,Roura2021,DiPumpo2021,DiPumpo2023}, measurements of fundamental constants \cite{Safronova18,Morel2020,Parker2018,doi:10.1126/science.1135459,Rosi2014}, realization of an optical mass reference \cite{Lan2013}, tests of contemporary dark matter \cite{PhysRevD.97.075020} and dark energy \cite{ChameleonDarkEnergy,Jaffe2017} candidates, and tests of theories of modified gravity \cite{QTEST,PhysRevLett.100.031101}.

On Earth, atom interferometers of various designs have been studied to capitalize on the favorable quadratic scaling of inertial-force sensitivities with matter-wave evolution times between the interferometry light pulses. Notably, high-precision measurements of UFF have been achieved with atom interferometers utilizing 10-meter-tall atomic fountains operating in Stanford University \cite{PhysRevLett.125.191101} and the Wuhan Institute of Physics and Mathematics \cite{Zho15}. Further, long-baseline interferometers utilizing alkaline earth metals are under construction at the Leibniz University of Hanover \cite{Hartwig_2015} and Stanford University~\cite{Abe2021} for precision fundamental physics experiments. Quantum gas experiments at the 100-meter drop tower in Bremen Germany have provided an enabling platform for developing cooling protocols and demonstrating atom interferometry in microgravity \cite{PhysRevLett.127.100401,doi:10.1126/science.1189164,PhysRevLett.110.093602}. The Einstein Elevator and the Gravi-Tower Pro expand this capability with greater than 100 drops per day providing 4~seconds and 2.5~seconds of freefall, respectively \cite{Einstein_Elevator,GraviTower}. Experiments have also been conducted in suborbital parabolic flights to detect inertial effects \cite{Geiger11} and to test the UFF with a dual-species atom interferometer in microgravity \cite{Barrett16}. Finally, the 100-meter-long MAGIS interferometer~\cite{Abe2021} and the AION interferometer network~\cite{Badurina_2020} are being developed for gravitational wave detection. Such work is expected to continue in multiple terrestrial facilities dedicated to the fundamental study of ultracold gases in microgravity conditions.

Developing matter-wave interferometry to operate in a space environment can address similar goals with significant advantages, including: (i) access to essentially unlimited freefall time in a compact instrument, (ii) enabling  novel schemes such as extreme adabatic cooling and delta-kick collimation for achieving ultra-low effective temperatures in atomic gases~\cite{Aveline2020, Pollard2020, Gaaloul2022}, and (iii) implementation and maturation of this quantum technology on a flight platform to support upcoming space missions where high-precision inertial sensing will be needed, including Earth observers and missions to test fundamental physics  \cite{QTEST, Frye2021, Chiow2018, ahlers2022stequest, Lachmann2021}. Even for trapped atom interferometers, which have demonstrated unprecedented coherence times for interferometry and provide localized gravity measurements \cite{Xu2019}, microgravity can be beneficial since the need for a relatively strong potential to support the atoms against gravity is no longer a constraint. In 2017, the MAIUS sounding rocket mission provided a seminal demonstration of cold atom technologies by operating for 6 minutes in-space flight and achieved the production of BEC of $^{87}$Rb atoms along with Bragg splitting and matter-wave interferometry \cite{Becker2018,Lachmann2021}.

In this article, we report on the first PI-led experiments exploring atom interferometry on an Earth orbiting platform. Our results include (i) the demonstration of high-visibility Mach-Zehnder interferometry (MZI) at relatively short times, (ii) signatures of atom interferometer sensitivities and visibility degradation due to the ISS vibration environment, (iii) the use of shear-Ramsey atom interferometry to extract phase shifts in a single experimental run, (iv) observation of interference fringes persisting for greater than 150~ms during freefall in the compact CAL science cell, and (v) the use of the CAL atom interferometer to perform a photon recoil measurement to demonstrate its utility as the first quantum matter-wave sensor in space.

\section{Results}

\begin{figure*}
\centering
\includegraphics[width=\textwidth]{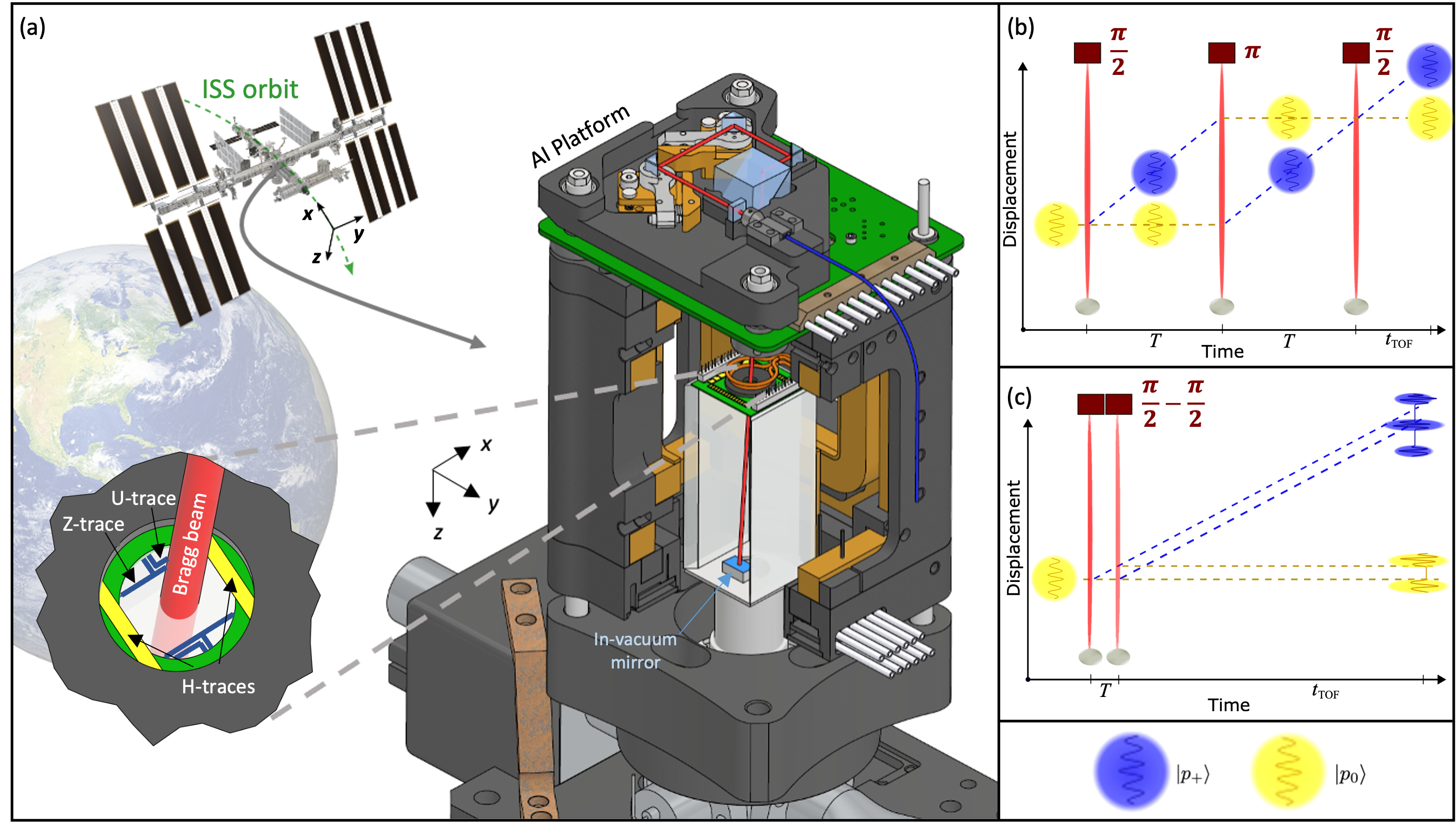}
\caption{Atom interferometer set up onboard the ISS. (a) Cut of the upper region of the physics package of SM-3 to expose the interior components and the path of the retro-reflected Bragg beam (red) inside the vacuum system. The expanded region shows the beam entering the vacuum chamber through a window and between pairs of Z and U-traces (blue) and H-traces (yellow) on the atom chip. (b) Space-time diagram of an ideal Mach-Zehnder interferometer (MZI), where three retro-reflected laser pulses (red) are applied in a sequence of $\frac{\pi}{2}$-$\pi$-$\frac{\pi}{2}$ pulses, with pulse separation times $T$, to create a superposition and then recombination of the initial atom cloud into two motional states ($\ket{p_+}$ and $\ket{p_0}$ given by blue and yellow clouds respectively). (c) Ramsey atom interferometer diagram using a sequence of $\frac{\pi}{2}$-$\frac{\pi}{2}$ pulses to explore shear-wave atom interferometry at long time-of-flight $t_\mathrm{TOF}$. For sufficiently short $T$, used for photon recoil measurements, the outputs are similar to that of the MZI.}
\label{fig:diagrams}
\end{figure*}

The CAL atom interferometer permits sensing of inertial forces onboard the ISS and is intended as a technology demonstrator for future space-borne fundamental physics experiments. Figure \ref{fig:diagrams}(a) gives an overview of the interior layout within the physics package at the heart of SM-3. This atom-interferometer-capable science module is based on the original CAL design \cite{Aveline2020,CALug}. Primary upgrades are a redesigned atom chip and the inclusion of a laser beam, which propagates from the atom interferometer platform above the science cell, through the center of a 3-mm diameter on-chip window (and through a set of atom-chip wires separated by 2~mm), and is retro-reflected from an in-vacuum mirror at the bottom of the science cell.  The CAL atom interferometer utilizes Bragg diffraction laser pulses \cite{Altin_2013,PhysRevLett.100.180405} operating at 785~nm for matter-wave beam splitters and mirrors. The Bragg beam waist is approximately 0.5~mm ($1/e^2$) within the science cell, and propagates at a 4$^{\circ}$ angle off the normal to the atom chip such that the laser wave vector $\bm{k}$ is nominally aligned with Earth's gravity vector along $\bm{z}$. Specifics of the design and implementation of the CAL atom interferometer were given previously \cite{Elliott2023}, with additional details in Methods section (\ref{Sect:MethodsAIDesign}).

For all experiments described in this article, ultracold gases of bosonic $^{87}$Rb are produced in the flight system using an atom-chip-based BEC source, producing up to 10$^4$ degenerate atoms approximately once per minute \cite{Aveline2020,CALug,Elliott2023}. Thereafter, a multi-stage transport protocol is used to quasi-adiabatically displace the magnetic trap minimum to position the atoms near the center of the Bragg beam, approximately 1~mm below the on-chip window. After transport, atoms in the $\ket{F=2, m_F = 2}$ internal state are confined in a nearly harmonic potential with measured trap frequencies ($\omega_x, \omega_y, \omega_z) = 2\pi\times (13.8, 23.4, 18.8)$ Hz, characterized using dual-axis absorption imaging. Atoms rapidly released from this center trap serve as the starting point for the following atom-interferometer experiments.

\subsection{Mach-Zehnder Interferometer}\label{Sect:DAIEO}
A three-pulse MZI \cite{Kasevich1992}, illustrated in Figure \ref{fig:diagrams}(b), 
was first demonstrated in Earth's orbit including eight repeated measurement campaigns over a span of 38 days. Here, $^{87}$Rb atoms were prepared and released from the center trap such that the cloud expands to less than 100~{\textmu}m FWHM over 100~ms of freefall, and moves with the center-of-mass velocity $\bm{v} = \bm{p}/m = (v_x, v_y, v_z) = (2.0,  1.0, 7.8)$~mm/s. The relatively large initial velocity ($p_0/m=v_0 \approx v_z$) along the laser wave vector $\bm{k}$ was intentionally applied so that the Bragg transitions to $\ket{p_{\pm}} \equiv \ket{p_0 \pm 2\hbar k} $ are nondegenerate, as this choice simplifies the interferometer operation and suppresses double diffraction \cite{Hartmann2020} (see Methods \ref{Sect:MethodsBraggAI} for details on the relevant Bragg processes). For this demonstration, the Bragg laser contains two frequencies separated by $\delta = 34.82$~kHz to compensate for the Doppler shift of the $^{87}$Rb atoms along the direction of the beam. During the first 17~ms of freefall, atoms were transferred to the magnetically-insensitive $\ket{F=2, m_F = 0}$ state via Adiabatic Rapid Passage, leaving no atoms detected in any of the unwanted $m_F \neq 0$ levels (see Methods \ref{Sect:MethodsAtomSource}). Thereafter, three square pulses of the Bragg laser approximating ($\pi/2,\pi,\pi/2$) transitions were applied for (0.13~ms, 0.25~ms, 0.13~ms), respectively, with $T = 0.5$~ms pulse separation times. 

\begin{figure}
\centering\includegraphics[width=\linewidth]{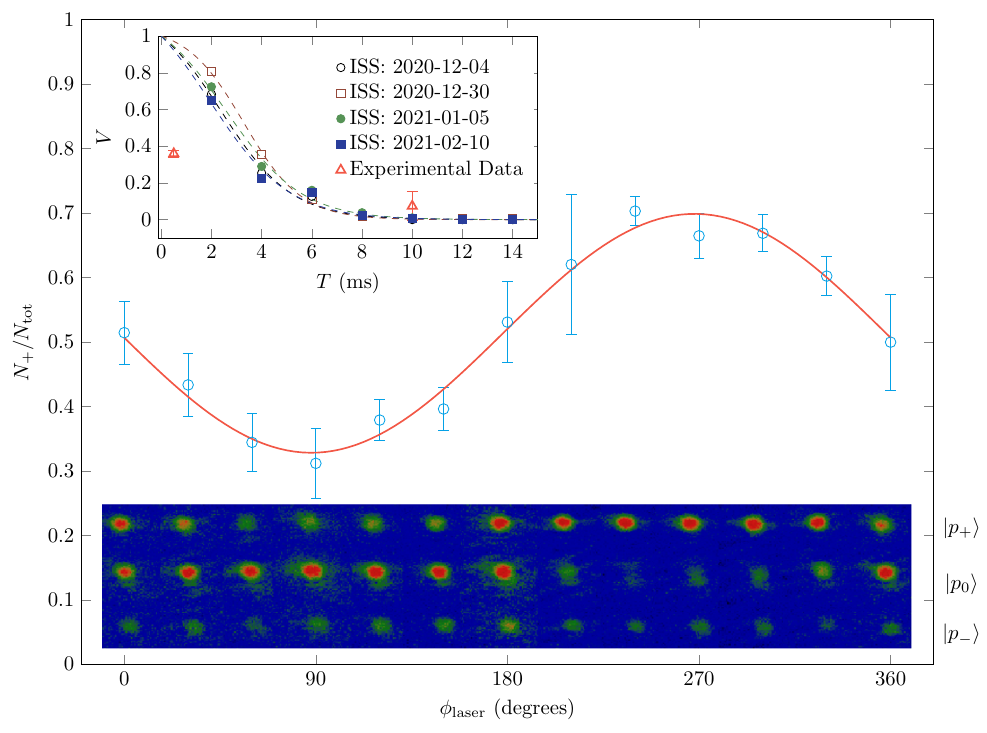}
\caption{Mach-Zehnder interferometry and the influence of vibrations on the ISS. The main plot shows the relative population of rubidium atoms in the $\ket{p_+}$ state after a Mach-Zehnder pulse sequence with $T = 0.5$~ms duration between Bragg pulses. Scanning the phase $\phi_\mathrm{laser}$ of the traveling wave for the final laser pulse reveals a corresponding sinusoidal variation of $N_+/N_\mathrm{tot}$. Eight independent data sets were analyzed, with up-to eight repetitions for each set. To increase the signal-to-noise ratio, the repeated images in each phase set were first summed, and the corresponding averaged images were fit to Thomas-Fermi profiles as described in Methods \ref{Sect:MethodsDetection}. The data plotted in light blue then yield the average $N_+/N_\mathrm{tot}$ for each $\phi_\mathrm{laser}$, with error bars given by the standard deviations. (Lower Inset) The first set of averaged absorption images after MZI, showing atoms oscillating between the $\ket{p_+}$ and $\ket{p_0}$ momentum states as $\phi_\mathrm{laser}$ is increased. A small population occupying the $\ket{p_-}$ state can also be seen. (Upper Inset) The influence of ISS vibrations is modeled (see Methods \ref{Sect:MethodsVibrations}), with 2~ms granularity, to illustrate limitations to the atom-interferometer visibility for single-source $^{87}$Rb BECs at larger $T$. Data for the acceleration $a_z$ in the $\bm{z}$-direction from the SAMS 121F04 accelerometer on the ISS was used for each day during which the MZI experiments were conducted, with dashed lines to guide the eye. Experimental results for $T = 0.5$~ms and 10~ms are included (orange triangles) for comparison.}
\label{fig:diagrams2}
\end{figure}

The phase difference $\phi$ between the two paths of a MZI is sensitive to external perturbations on the atoms, including accelerations $\bm{a}$, rotations $\bm{\Omega}$, and the Bragg laser phase \cite{Bongs2006,Borde2004,Hogan2008}. 
In leading order, $\phi$ can be expressed by:
\begin{equation*}
  \phi = \big[\bm{k_\mathrm{eff}} \cdot \bm{a} + 2(\bm{k_\mathrm{eff}} \times \bm{v})\cdot \bm{\Omega}\big] T^2 + \phi_\mathrm{laser},
\end{equation*}
where $\bm{k_\mathrm{eff}}$ is the effective two-photon wave vector with amplitude $\approx 2{k}$ that defines the transferred momentum during diffraction. To map out an interference fringe, the differential phase of two simultaneous frequency tones in the final Bragg laser pulse $\phi_\mathrm{laser}$ was varied between 0$^\circ$ and 360$^\circ$. 
After the final Bragg pulse, the atoms were allowed to propagate freely for $t_\mathrm{TOF}$ = 15~ms so that the different momentum components, constituting the exit ports of the atom interferometer, could separate in position. The populations of atoms $N_0$ and $N_+$, occupying the $\ket{p_0}$ and $\ket{p_{+}}$ output momentum states, respectively, were then measured using absorption imaging as shown in the lower inset of  Figure \ref{fig:diagrams2}. Note that a small population in the spurious exit port $N_-$, corresponding to atoms driven towards the atom chip via off-resonant transitions into the $\ket{p_-}$ state, was also produced as a result of the finite cloud temperature and moderate two-frequency detunings of the Bragg laser \cite{Hartmann2020,Becker2018,Lachmann2021}. Figure \ref{fig:diagrams2} shows the characteristic sinusoidal variation of the excitation fraction $N_+/N_\mathrm{tot}$ with $\phi_\mathrm{laser}$ (see Methods \ref{Sect:MethodsBraggAI} Eq.~\eqref{Eqn:fringe_fit}), providing a clear and repeatable signature of matter-wave interference with visibility $V = 0.36(2)$.

The limited visibility of the interferometer with short $T$ can be attributed to two effects: (i) Condensates released from the trap expand due to mean-field interactions, leading to a Thomas-Fermi velocity spread along the $\bm{k}$-direction of $\Delta v_0 \approx 1.6$~mm/s. Both the Bragg pulse efficiencies and the phases developed during free evolution depend on the atomic velocity. (ii) Analysis of the Bragg pulse efficiencies suggests that the atoms experience a relative spread in Bragg beam intensity of approximately $\pm 25$\%. This variation is significantly larger than the approximately 5\% variation expected from the spatial inhomogeneity of the Bragg beam if it had an ideal Gaussian profile. However, optical simulations indicate that the collimation lens used in the Bragg beam fiber coupler produces an imperfect Gaussian with substantial wings at large radii. The wings are clipped by the 2-millimeter aperture given by the atom-chip traces (see Methods \ref{Sect:MethodsAIDesign}) and the resulting diffraction pattern produces significant intensity modulation. We confirmed this effect by applying long-duration pulses of Bragg light to a $^{87}$Rb BEC and observing the transverse motional excitation produced by the BEC propagating through this diffraction pattern.

Increasing $T$ from 0.5~ms to 10~ms causes the interferometer visibility to drop to a value consistent with zero. However, the shot-to-shot signal variations are on par with the visibility observed at $T = 0.5$~ms, signifying that phase noise is dominating the signal \cite{Gersemann2020}. This behavior is consistent with the simulated $T$-dependent visibility degradation, shown in the upper inset of Figure \ref{fig:diagrams2}, which was modeled using ISS-SAMS 121F04 accelerometer data measured along the $\bm{z}$ direction. For details of the visibility degradation simulation, see Methods \ref{Sect:MethodsVibrations}.

\subsection{Shear-Wave Atom Interferometry}\label{Sect:ShearWaveAI}

In Figure \ref{fig:diagrams}(c) a two-pulse Ramsey interferometer geometry \cite{RevModPhys.81.1051} is illustrated. Here, two $\pi/2$ pulses of the Bragg laser beam are applied in relatively rapid succession so that the atomic wave functions split from the first laser pulse still sufficiently overlap as to interfere. Depending on the expansion energy of the atoms and the freefall time after the second pulse $t_\mathrm{TOF}$, as well as the timings of the interferometer sequence and $\phi_\mathrm{laser}$, such shear-wave interferometers can provide direct information about both the atoms and the experimental apparatus.

\begin{figure}
\centering\includegraphics[width=\linewidth]{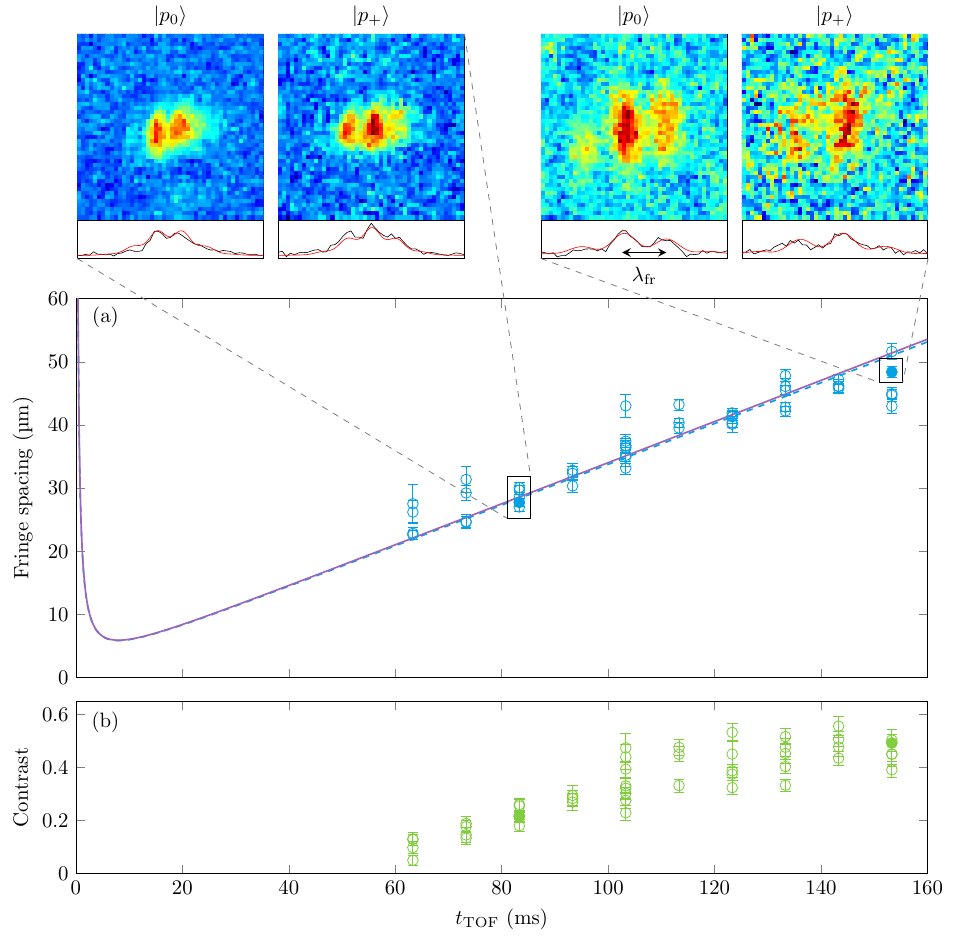}
\caption{Single shot shear-Ramsey interference in orbit. 
(a) Fringe spacing $\lambda_{\mathrm{fr}}$ of a shear-Ramsey interferometer aboard CAL as a function of the free-evolution time $t_{\mathrm{TOF}}$ of the BEC after release from the magnetic trap. The BEC is split coherently with two $\pi$/2 Bragg pulses each lasting 0.16 ms with a separation of 1.0~ms between them to achieve the characteristic interference pattern for both exit ports within a single experimental run. The measured fringe spacings (blue dots for data and dashed line for the fit) show excellent agreement with the theoretical prediction (purple line) based on the expansion dynamics of the BEC in the Thomas-Fermi regime and a proper treatment of the finite pulse duration as expressed by Eq.~\ref{eq:fringe_spacing}.
Exemplary 2D-density distributions for two different expansion times are shown in the top row ($189 \times 189$~\textmu $\mathrm{m}^2$ area, with densities normalized for each image individually) with the integrated 1D densities (black lines) displayed below together with the fit results (red lines). (b) The contrast of the interference fringes (green dots) increases with expansion time and saturates at a maximum of around 45\%. Error bars represent 1-sigma confidence bounds of fitting the density distributions.}
\label{fig:diagrams3}
\end{figure}

For this interferometer arrangement, sufficient $t_\mathrm{TOF}$ is implemented after the pulse sequence so that spatial fringes are visible in a single experimental shot, as shown in Figure~\ref{fig:diagrams3}. The fringes persist for over 150~ms time of flight, which would be challenging to observe in a similarly compact terrestrial instrument. Here, two $\pi/2$ Bragg laser pulses, each with a square-pulse time distribution lasting 0.16~ms and separated by 1~ms, are applied 2 ms after release of the BEC from the trap. Approximately 60~ms after the second pulse, weak stripe patterns emerge in both the undiffracted cloud ($\ket{p_0}$ state) and the cloud that was diffracted by the Bragg beam towards the atom chip (defined here as the $\ket{p_+}$ state). A weak third cloud, not shown in Figure~\ref{fig:diagrams3}, consists of atoms off-resonantly excited to the $\ket{p_-}$ state. The images cannot be reliably analyzed for times greater than 150~ms because atoms in the $\ket{p_+}$ state propagate into the region near the chip which exhibits degraded signal quality and, eventually the ultracold cloud is destroyed as the atoms come into contact with the chip itself. Due to resolution limits of the detection system and the size of the BEC being smaller than the fringe spacing for very short times, the nonlinear behavior of the fringe spacing shown in Figure~\ref{fig:diagrams3} could not be resolved. However, there is promise in exploring this nonlinear regime with greater resolution in the future.

Fits to extract the contrast and fringe spacing for the two primary clouds were obtained by first integrating the density distribution of the clouds along a direction parallel to the stripes, and then fitting each cloud to a Gaussian distribution with sinusoidal oscillations and a constant phase difference of $\pi$. 
Figure~\ref{fig:diagrams3} gives the measured fringe spacing $\lambda_{\mathrm{fr}}$ and contrast as a function of $t_\mathrm{TOF}$. Each point represents data from a single experimental run with error bars given by the fit uncertainty. A theoretical prediction for the fringe spacing (purple line) with no adjustable parameters shows excellent agreement with the fringe spacing data. The model, which takes into account the expansion dynamics of a BEC in the time-dependent Thomas-Fermi approximation, gives the following result for the fringe spacing \cite{Roura2014}:
\begin{equation}\label{eq:fringe_spacing}
    \lambda_{\mathrm{fr}}(t_\mathrm{TOF}) = \frac{\pi}{k} \frac{R(t_\mathrm{TOF})}{T_{\mathrm{eff}}\dot{R}(t_\mathrm{TOF})}
\end{equation}
Here, $R(t_\mathrm{TOF})$ is the Thomas-Fermi radius, which follows from the scaling approach~\cite{Meister2017}. The asymptotic constant expansion rate is determined by the initial trap frequencies and the number of atoms and $T_{\mathrm{eff}}$, the effective time separation of the two Bragg pulses taking into account the finite duration of each pulse. We calculated $T_{\mathrm{eff}}$ by solving the Schr\"{o}dinger equation in the presence of the Bragg optical potential, giving a value of $T_{\mathrm{eff}} = 1.204$~ms. When instead we fit the data to determine $T_{\mathrm{eff}}$, we obtain a value of 1.21(1)~ms, in excellent agreement with the calculated value. 

Because the shear-Ramsey interferometer yields an interference pattern in a single run, it is \emph{not} sensitive to dephasing by vibrational phase noise. 
The ability to observe the pattern after long freefall times illustrates the potential for exploring long-time interference effects in microgravity.
A similar terrestrial experiment, limited to expansion times of a few milli\-seconds, illustrated the effects of atomic interactions on the fringe patterns~\cite{Simsarian_2000}. For our weak initial trap and low atom numbers, however, interactions between the two clouds can be safely neglected.

\subsection{Photon Recoil Measurement}\label{Sect:PhotonRecoil}

The two-pulse Ramsey interferometer configuration described in Section \ref{Sect:ShearWaveAI} can also be used to perform a proof-of-principle recoil frequency measurement. Notably, in between the two Bragg pulses, the atomic wave packets develop a phase difference that depends on the recoil frequency $\omega_r \equiv \hbar k^2/2m$. High-precision terrestrial measurements of $\omega_r$ are used to determine the fine structure constant \cite{Parker2018, Morel2020}. 
The long interrogation times available in microgravity could support such efforts, and even a lower-precision recoil frequency measurement could provide a useful method to calibrate the Bragg wave number $k$ in a space environment where no conventional wave\-meter instrument is available.

The phase evolution of a free atom is governed by its kinetic energy $p^2/2m$. 
The interferometer is sensitive to the phase difference between wave packets with momenta $p_\pm$ and $p_0$, which results in $\phi_\mathrm{atom} = -(4\omega_r \pm 2 k v_0)T$ for pulse separation time $T$ (see Methods Eq.~\eqref{eq:QS2}). In addition, the laser phase evolves as $\delta T + \phi_0$, where $\delta$ is the difference between frequencies in the Bragg laser and $\phi_0$ is the phase offset of the second Bragg pulse. The measured phase, from Methods Eq.~\eqref{Eqn:phi}, is therefore:
\begin{equation*}
  \phi = -(4\omega_r \pm 2 k v_0)T + \delta T + \phi_0.
\end{equation*}
In order to interferometrically determine both $\omega_r$ and $v_0$, we measure $\phi$ as a function of $T$ for both signs of the Bragg momentum kick. 

\begin{figure}
\centering\includegraphics[width=\linewidth]{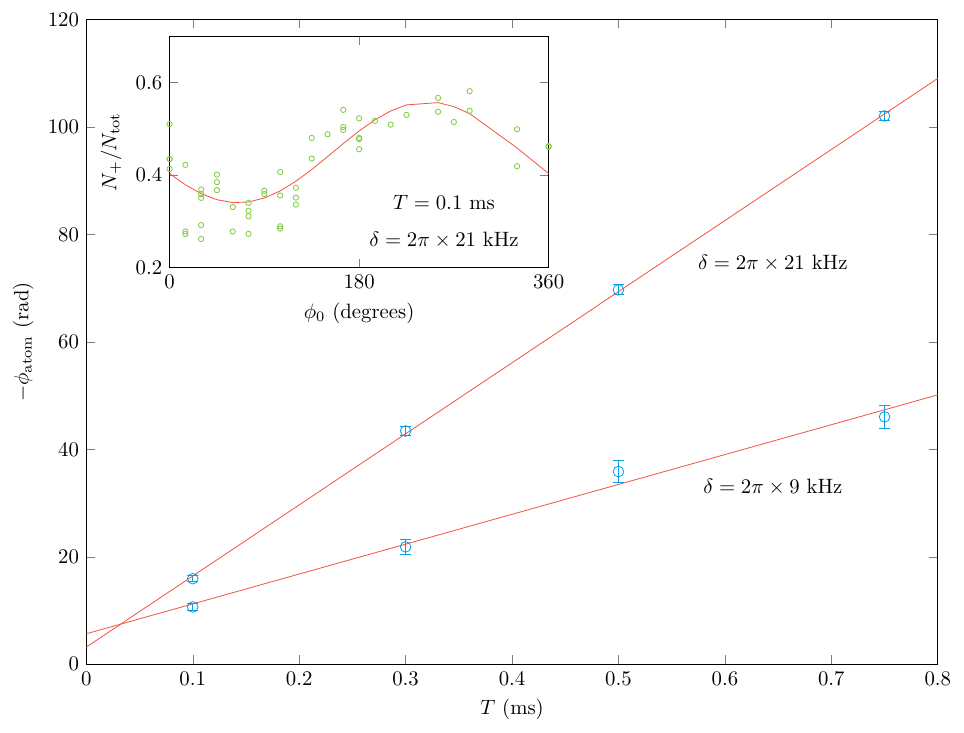}
\caption{Recoil frequency measurement in a Ramsey interferometer. Accumulated phase difference of the atomic wave packets as a function of interrogation time $T$. 
Circled points are obtained by fitting interference fringes as shown in the inset, and error bars represent the estimated fit uncertainties. The lines are linear fits for the Bragg transition $\ket{p_0} \rightarrow \ket{p_+}$ with frequency $\delta = 2\pi\times 21$~kHz and $\ket{p_0} \rightarrow \ket{p_-}$ with  $2\pi\times 9$~kHz. The phases for the two cases differ at $T = 0$ because of different couplings to their respective off-resonant momentum states.}
\label{fig:Recoil}
\end{figure}

In these experiments $v_0 \approx 2.5$~mm/s, so for the $\ket{p_0} \rightarrow \ket{p_+}$ transition away from the atom chip we set $\delta = 2\pi\times 21$~kHz and for $\ket{p_0} \rightarrow \ket{p_-}$ we set $\delta = 2\pi\times 9$~kHz. The net phase was determined by scanning $\phi_0$, as shown in the inset of Figure~\ref{fig:Recoil}. The data is fit to a sinusoidal function to extract $\phi_\mathrm{atom}$ at various values of $T$, with results shown in the main plot of Figure~\ref{fig:Recoil}. From the slopes of these lines, we find $\omega_r = 2\pi\times 3.77(6)$~kHz and $v_0 = 2.4(1)$~mm/s. The recoil frequency is consistent with the expected value of 3.72~kHz, and the initial velocity is consistent with time-of-flight measurements.

The precision of these measurements is low due to the limited interrogation time. 
At the maximum usable time of $T = 0.75$~ms, the interference visibility has dropped to about 0.07 which is consistent with the expected coherence time $R_0/v_B \approx 0.7$~ms. Here, $R_0 \approx 8$~\textmu{}m is the initial condensate size in the trap
and $v_B \equiv 2\hbar k/m = 11.7$~mm/s denotes the Bragg recoil velocity. Longer interaction times could be achieved by starting with atoms in a weaker trap, but a more effective method would be to use a closed interferometer configuration such a three-pulse contrast interferometer~\cite{PhysRevLett.89.140401}.

\section{Discussion}

In this article, we have reported on three unique experiments based on matter-wave interference of ultracold rubidium gases in orbital microgravity, which were each found to be robust and repeatable over timescales spanning months. We utilized the long freefall times available for atoms in the relatively compact CAL apparatus and compared the atom interferometer performance with ISS environmental data to gain insight into the high-frequency vibration environment experienced by the instrument. We further employed shear-wave interferometry to demonstrate the effects of matter-wave interference that are clearly visible in a single experimental run for more than 150~ms of freefall. Additionally, photon recoil measurements served to provide characterizations of the Bragg wave number using only the atomic feedback resulting from interferometer sequences, demonstrating the utility of CAL as the first quantum matter-wave sensor in space.

Significant advancements and new sensing capabilities are expected to arise as differential interferometry (using either spatially-separated atom interferometers with the same atomic species or simultaneous interrogation of two distinct atomic species) is further explored with CAL~\cite{Gersemann2020, Foster:02, Varoquaux_2009}. 
The CAL atom interferometer is designed for precision dual-species simultaneous interferometry with $^{87}$Rb and $^{39}$K or $^{41}$K, interrogated via absorption imaging for phase-sensitive measurements with two dissimilar quantum test masses. Specifically, the operating frequency of the Bragg laser at 785~nm is near the ``magic" wavelength where the two-photon Rabi frequencies are essentially identical for Rb and K species \cite{QTEST}. Hence, differential atom interferometer measurements in CAL are expected to be minimally impacted by common-mode noise sources including vibrations and laser noise. Initial experiments \cite{Elliott2023} on CAL demonstrating differential atom interferometry with $^{87}$Rb and $^{41}$K quantum gases at short $T$ are particularly promising for PI-led efforts aiming to extend the capabilities of space-based dual-species differential atom interferometry.

A replacement for SM-3 on the ISS is already designed for reduced scattering of the beam before it enters the science cell and, hence, minimizing wavefront degradation. 
As a CAL follow-on, the BECCAL mission~\cite{Frye2021} is also expected to include numerous mechanisms to optimize visibility and allow long interrogation times with differential atom interferometers, including using relatively large beams and employing rotation compensation schemes to counter the effects of the ISS rotation as it orbits the Earth.  

The experiments detailed in this article therefore serve as pathfinders for proposed space missions relying on sustained matter-wave interferometry with unprecedented sensitivity to inertial and fundamental physical forces, including tests of 
Einstein's Equivalence Principle~\cite{QTEST,ahlers2022stequest,tarallo2014}, 
gravitational wave detectors~\cite{Kolkowitz16,PhysRevA.94.033632,AGIS_LEO,Yu11,Badurina_2020,Abe2021,Canuel2018}, 
direct detection of dark matter and dark energy candidates \cite{ChameleonDarkEnergy,QTEST}, 
Earth and planetary sciences including geodesy, seismology, and sub-surface mapping~\cite{Yu2006, Sorrentino2011, Chiow2018}, and 
advanced navigation and drag-free referencing \cite{Battelier2016,Fang2016}. 
With the successful utilization of CAL SM-3 for the first long-term PI-led studies of atom interferometry in Earth's orbit, this first-of-its kind instrument has brought quantum sensing via atom interferometry into the list of technologies matured and flight-qualified by NASA towards expanding our fundamental understanding and exploration of the cosmos.

\section{Methods}

\subsection{Upgraded Cold Atom Lab Facility}\label{Sect:MethodsAIDesign}
The CAL atom interferometer is designed to be sensitive to the force of Earth's gravity, rotations of the ISS, and also to utilize differential measurements from atom interferometers consisting of simultaneously interrogated rubidium and potassium gases. To this end, the on-orbit upgraded SM-3 includes four primary changes from the original CAL design to enable atom interferometry \cite{Elliott2023}, the first three of which are illustrated in Figure \ref{fig:diagrams}(a).  (i) Light from the Bragg laser system was delivered into the SM-3 via polarization maintaining optical fibers and routed to the atom interferometer platform, which is the optical bench used to shape and direct the beam as part of the upgraded physics package provided by ColdQuanta. (ii) The atom chip was upgraded to allow the beam to propagate through the center of the 3-mm diameter window on the atom chip, which also serves as the top surface of the evacuated science cell. Passing the beam through the chip, as opposed to passing through the cell windows horizontal to the chip, was necessary to accommodate the desired orientation of the Bragg beam with respect to gravity. Providing clearance for the laser required the atom chip traces to be moved away from the center of the window, resulting in two pairs of in-vacuum traces (Z- and U-traces respectively), separated by approximately 2~mm along the $x$-axis, and a pair of wires on the atmospheric side of the chip (H-traces), separated by approximately 3~mm and oriented along the $y$-axis. (iii) An in-vacuum mirror was mounted in the science cell on the opposite side of the atom chip, slightly off center and oriented at a 4$^{\circ}$ angle to allow retro-reflection of the Bragg beam while assuring clearance for the push beam and atom flux from the 2D MOT \cite{Aveline2020}. (iv) The optical path for the through-chip imaging system was rerouted to accommodate detection through the chip window in combination with the atom interferometer platform. 

The fiber-coupled Bragg laser was included as part of the original CAL flight system in anticipation of SM-3. Coherent light is provided by a single-wavelength external cavity diode laser operating at nominally 785~nm, which exhibits a narrow linewidth ($<$~200 kHz) and continuous output of tens of milliwatts. The light is directed via optical fibers from the single locker to the quad-locker \cite{Aveline2020} where it is amplified using a Tapered Amplifier (TA). The TA input is shared between the 785~nm and 780~nm seed lasers and the output is directed either to the cooling light or atom interferometer beam paths via optical switches. Before the 785-nm light is directed into SM-3, it is passed through an acousto-optic modulator, operating at approximately 79~MHz as controlled by an arbitrary waveform generator. This design allows multiple frequency components to be written onto the Bragg laser.  A narrow bandpass optical filter (Semrock Laser Clean-up MaxLine 785/3) on the atom interferometer platform passes 785-nm light while suppressing residual light near 780~nm by over three orders of magnitude, in order to avoid resonant scattering. The resultant Bragg beam was approximately collimated within the CAL science cell and retro\-reflected by the in-vacuum mirror with overlap of the incident and reflected beam, measured before launch to the ISS, better than 0.1~mm. 

\subsection{Source Control and Detection} \label{Sect:MethodsAtomSource}
 
Transfer of $^{87}$Rb atoms from the magnetically sensitive $\ket{F=2,m_F=2}$ state (within the $^{2}S_{1/2}$ manifold) to the magnetically insensitive $\ket{2, 0}$ state is achieved with high-field adiabatic rapid passage (ARP) \cite{KrutzikPhD}. Here, the second-order Zeeman shift is made to be much larger than the Rabi frequency for RF coupling of atoms among magnetic sublevels, breaking the degeneracy that would otherwise arise when attempting to drive atoms among the five $m_F$-levels in the $F=2$ manifold. At an applied bias field of 29.2~G, the splitting between the sublevels is approximately 20 MHz, and the second-order Zeeman shift is 120 kHz. In comparison, the RF Rabi frequency for the $\ket{2,2} \rightarrow \ket{2,1}$ transitions is approximately 9 kHz. The atoms are initially released from the center-trap and allowed to freefall for 5.2~ms while a $B_x \approx 31$~G bias field along the x-axis is applied and allowed to stabilize. Single-tone RF at 20.4 MHz is then pulsed on for 5~ms while $B_x$ is linearly ramped down to approximately 29.2~G. Using this method, an ultracold rubidium gas is transferred from the $\ket{2, 2}$ to the $\ket{2, 0}$ state with no detectable signal remaining in other $m_F$ states.  Thereafter, $B_x$ was turned off and a 10 mG magnetic field was applied along the $\bm{y}$ direction to maintain a field quantization axis throughout the freefall and atom interferometer stages. 

Atomic density distributions are measured via one of two absorption imaging systems, which are nearly identical to those of the original CAL science module \cite{Aveline2020}, aside from the change to the through-chip imaging path discussed above. Here, the ``horizontal'' imaging system is used for detection of interference as it images the $x$-$z$ plane. The ``vertical'' imaging system observes the atoms in the $x$-$y$ plane through the atom chip and was used to help position atoms in the Bragg beam. At short atom interferometer interrogation times, the average signal variation at a fixed phase is $\sigma = 0.06$, approximately an order of magnitude larger than the expected quantum projection-noise limit \cite{PhysRevA.47.3554} for an ensemble of $N_\mathrm{tot} \sim 5 \times 10^3$ atoms. The excess can be explained by detection noise in the absorption imaging technique.

\subsection{Bragg Atom Interferometry}\label{Sect:MethodsBraggAI}

The principles of light-pulse atom interferometry have been reviewed in detail elsewhere~\cite{Kasevich1992, RevModPhys.81.1051,Kleinert2015}. Here, we summarize the related theory for understanding the phases-shifts and fringes observed in Figures~\ref{fig:diagrams2}, \ref{fig:diagrams3}, and \ref{fig:Recoil}. For simplicity, we consider the ideal first-order Bragg process driven by a light pulse with two frequency components $\omega_1$ and $\omega_2$, counter-propagating 
wave vectors $\bm{k_1}$ and $\bm{k_2}$, and frequency difference $\delta \equiv \omega_1 - \omega_2$. For small detunings relevant to this study, $k_1 \simeq k_2 = k$. The Bragg light pulse coherently couples atoms occupying an initial momentum state $\ket{p_0}$ to the final states $\ket{p_{\pm}} \equiv \ket{p_0 \pm 2\hbar k} $. Neglecting additional inertial forces, the evolution of the quantum state $\ket{\Psi(t)}$ can be expressed in terms of time-dependent coefficients $c_0$ and $c_\pm$ by
\begin{eqnarray*}
    \ket{\Psi(t)} &=& 
    c_0(t)\ket{p_0}\exp{\left[-\frac{i}{\hbar}\frac{p_0^2}{2m}t\right]} + \\ 
    &&  c_{\pm}(t)\ket{p_{\pm}} \exp{\left[-\frac{i}{\hbar}\frac{(p_0 \pm 2\hbar k)^2}{2m}t\right]}
\end{eqnarray*}
which, up to an overall phase, is equivalent to
\begin{equation}\label{eq:QS2}
    \ket{\Psi(t)} = c_0(t)\ket{p_0} + c_{\pm}(t)\ket{p_{\pm}}e^{-i(4\omega_r \pm 2 k v_0) t},
\end{equation}
with recoil frequency $\omega_r \equiv \hbar k^2/2m$ and initial velocity $v_0 = p_0/m$ along $\bm{k}$. If the two levels are coupled by a pulse with duration $\tau$, Rabi frequency $\Omega$, and frequency $\delta$, then the detuning is $\Delta_{\pm} = 4\omega_r \pm 2 k v_0 - \delta$ and the probability of finding an atom in the $\ket{p_{\pm}}$ state reads
\begin{equation*}
    |c_{\pm}(\tau)|^2 = \frac{\Omega^2}{\Omega^2+\Delta_{\pm}^2}\left[ \sin{\left(\frac{\sqrt{\Omega^2+\Delta_{\pm}^2}}{2}\tau\right)} \right]^2.
\end{equation*}
For the resonant case $\Delta_{\pm} =0$, an equal superposition of $\ket{p_0}$ and $\ket{p_{\pm}}$ states is prepared for a $\pi/2$ pulse, where $\Omega \tau = \pi/2$ and atoms are fully transferred between states for a $\pi$ pulse, when $\Omega \tau = \pi$. 

Matter-wave interference can be observed by applying a series of $\pi/2$ and $\pi$ pulses, with pulse separation time $T$, as illustrated in Fig. \ref{fig:diagrams}. For each geometry, the output phase of the atom interferometer is observable from the population of atoms ($N_0$ or $N_+$) occupying one of two final momentum states ($\ket{p_0}$ or $\ket{p_+}$, respectively) as:
\begin{equation} \label{Eqn:fringe_fit}
    \frac{N_0}{N_{\mathrm{tot}}} = \frac{N_{\mathrm{mean}}}{N_{\mathrm{tot}}}(1+V\cos \phi)
\end{equation}
where $N_0 + N_+ = N_{\mathrm{tot}}$, $N_\mathrm{mean}$ is the number of atoms occupying $N_0$ averaged over all interferometer phases $\phi$, and $V$ gives the atom interferometer visibility. Therefore, observing the fraction of atoms in each final momentum state allows us to extract the accumulated phase modulo 2$\pi$. The measured phase can be decomposed into contributions from the Bragg laser and the atomic wave function evolution
\cite{Hogan2008}: 
\begin{equation}\label{Eqn:phi}
     \phi =  \phi_{\mathrm{laser}} +  \phi_{\mathrm{atom}}
\end{equation}
These contributions depend on the environment and forces experienced by the atoms, the frequency stability of the laser pulses, geometric topological phase shifts, and the geometry of the atom interferometer itself \cite{RevModPhys.81.1051}.

\subsection{Fringe Detection and Data Analysis}\label{Sect:MethodsDetection}

All interferometer data is acquired via absorption imaging, which yields the spatial distribution of the atoms at the end of the experimental sequence. Images with no atoms or other obvious errors are discarded. For the Mach-Zehnder and recoil measurements, images are cropped to a region of interest around the expected
position of each output cloud, and the clouds are fit to independent Thomas-Fermi profiles in order to extract the atom number. All fits are confirmed visually. For the shear Ramsey atom interferometer measurements, absorption images were first processed using principle component analysis \cite{Segal2010} to remove background imaging noise, and are then fit to the Ramsey fringe model described in the main text.

For the recoil frequency measurements, data sets of about 50 points are acquired by scanning the Bragg phase $\phi_0$ from 0 to $2\pi$. The resulting curves are fit to the form of Eq.~\eqref{Eqn:fringe_fit}, where here $\phi_\mathrm{atom} = -4(\omega_r\pm 2kv_0)T$ and $\phi_\mathrm{laser} = \delta T + \phi_0$. Uncertainty in $\phi$ is determined by the bounds for doubling the quality-of-fit parameter $\chi^2$. The uncertainty ranges from about 0.5~rad at small $T$ up to as much as 2 rad for large $T$, which is consistent with the reduction in atom interferometer visibility at larger $T$. In comparison, uncertainties in $\phi_0$, $\delta$ and $T$ are negligble.

\begin{figure*}
  \centering
  \includegraphics[width=\textwidth]{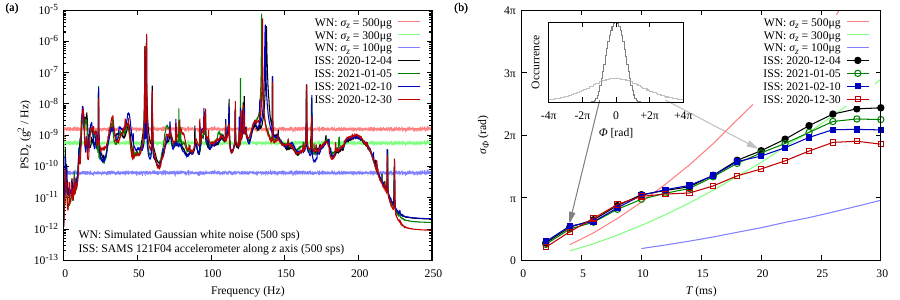}
  \caption{Influence of the ISS vibrations on the accumulated atom interferometer phase.
    (a) The spectra of the accelerations in $\bm{z}$ direction from the
    SAMS 121F04 accelerometer are compared to simulated Gaussian white noise spectra for different standard deviations $\sigma_z$. The accelerometer spectra are averaged over relevant time frames of $\unit[2\text{--}4]{h}$ for days during which atom interferometer experiments were conducted (see legend).
    (b) The standard deviation ($\sigma_{\Phi}$) of the distribution of accumulated phases is depicted as a function of $T$.
    Distributions were obtained by calculating and binning the accumulated phases after dividing the acceleration data in blocks corresponding to the respective times $T$. (inset) Two examples of these distributions are given for $T = \unit[4]{ms}$ and $T = \unit[20]{ms}$. Deviations of $\sigma_{\Phi}$ from the expected $T^2$-scaling for ISS data are attributed to deviations of the accelerometer spectra from that of ideal Gaussian white noise as shown.}
  \label{fig:ai:phase}
\end{figure*}

\subsection{ISS Vibrations Study}\label{Sect:MethodsVibrations}

Our analysis of the impact of the ISS vibrations on the atom interferometer visibility is based on $a_{z}$
data from the SAMS 121F04 three-axis accelerometer, which are provided publicly over
NASA's Principal Investigator Microgravity Services (PIMS)~\cite{McPherson2009}.  
The accelerometer is located on the front panel of the CAL Instrument in front of SM-3. 
It has a sample rate of $\unit[500]{sps}$ and a bandwidth of $\unit[200]{Hz}$.  
The Physics Package is attached via a flexure mount inside the science module which modifies the acceleration spectrum.  
However, we expect no major impact in the frequency range of interest and use the unmodified accelerometer data in our analysis for simplicity.

The SAMS accelerometer provides discrete measurements $a_{z,i}$ every $\Delta t = \unit[2]{ms}$, 
giving a total of $n = \lfloor 2T / \Delta t \rfloor$ measurements at our disposal for each Mach-Zehnder interferometer sequence. 
Equations for calculating the accumulated MZI phase from $a_z$, $k_\mathrm{eff}$ and the sensitivity function $g_\text{s}(t)$ \cite{Cheinet2008} were discretized to utilize the PIMS data sets. 
The durations of the Bragg laser pulses were ignored since they are short compared to the overall atom interferometer interrogation time. 
We obtain 
\begin{equation*}
  \Phi(T) \approx 
   k_\text{eff} \, (\Delta t)^2 \sum_{j = 1}^{2n} g_j \sum_{i = 1}^j a_{z,i},
\end{equation*}
with the discretized sensitivity function
\begin{equation*}
  g_j \equiv \begin{cases}
    -1 & \text{for } 1 \le j \le n\\
    +1 & \text{for } n < j \le 2 n
  \end{cases}.
\end{equation*}

Although it was not feasible to correlate accelerometer measurements to individual atom interferometer experiments, the
impact of vibrations was estimated for each day that MZI data was collected.
Specifically, the accelerometer data set from a representative time
frame of $\unit[2\text{--}4]{h}$ for each day was divided into blocks of $n$ measurements.
We then calculated the accumulated phase $\Phi(T)$ for each block at $T = \unit[2]{ms}, \unit[4]{ms}, \dotsc, \unit[30]{ms}$.
The phases follow Gaussian distributions (see Fig.~\ref{fig:ai:phase}(b), inset)
with standard deviations $\sigma_\Phi$ shown as a function of $T$ in the main frame of Fig.~\ref{fig:ai:phase}(b). 

The impact of the vibrational phase noise on the atom interferometer visibility was modeled using simulated data for an ideal fringe. For each $T$, Gaussian-distributed noise with the previously calculated $\sigma_\Phi$ was added to the  phase $\phi$ of ideal interferometer fringes. Fitting the resulting signals to Eq.~\eqref{Eqn:fringe_fit} yielded the expected visibility shown in the upper inset of Figure \ref{fig:diagrams2}.
While the ISS vibration spectra remained similar over the
roughly 3-month time span over which the MZI experiments were conducted,
differences in the day-to-day vibrations are noticeable at larger $T$.

For short $T$, $\sigma_\Phi$ due to ISS vibrations
is larger than $\sigma_\Phi$ from simulated Gaussian white noise.
However, the behavior inverts for larger $T$, as shown in Figure \ref{fig:ai:phase}(b).
A likely explanation for the deviation from the expected $T^2$-behavior is that the harmonics in the real ISS vibration
spectra are important for short $T$. The influence of small $n$ for $T = \unit[2]{ms}$ are also found to lead to slight deviations from an ideal Gaussian distribution even with simulated Gaussian white noise.
However, for larger $T$, low frequency harmonics are less impactful and can essentially only
contribute a maximum phase.

\section*{Acknowledgments}

We gratefully acknowledge the contributions of current and former members of CAL’s operations and technical teams, and those of the team in the ColdquantaLabs division of Infleqtion. 
We also recognize the continuing support of JPL's Astronomy, Physics, and Space Technology Directorate, of the JPL Communication, Tracking, and Radar Division, the JPL Mission Assurance Office, the Payloads Operations Integration Center (POIC) cadre at NASA’s Marshall Space Flight Center, the International Space Station Program Office (ISSPO) at NASA’s Johnson Space Center in Houston, and ISS crew members. 
We are thankful for dedicated support from the Biological and Physical Sciences Division (BPS) of NASA’s Science Mission Directorate at the agency’s headquarters in Washington, D.C. 

Cold Atom Lab was designed, managed, and operated by the Jet Propulsion Laboratory, California Institute of Technology, under contract with the National Aeronautics and Space Administration (Task Order 80NM0018F0581). 
CAL and the PI-led science teams, including J.R.W, C.A.S, D.C.A, S.B., E.R.E, J.M.K, H.M., K.O., L.P., M.S., C.Schn., B.S., R.J.T., and N.P.B., are sponsored by BPS of NASA’s Science Mission Directorate at the agency’s headquarters in Washington, D.C. and by ISSPO at NASA's Johnson Space Center in Houston. 
E.M.R., G.M., W.P.S., P.B., E.G., A.P., and N.G. acknowledge support by the DLR Space Administration with funds provided by the Federal Ministry for Economic Affairs and Climate Action (BMWK) under grant numbers DLR 50WM2245-A/B (CAL-II), 50WM2253A (AI-quadrat), 50WM2250E (QUANTUS+), and 50WM2177 (INTENTAS). 
N.G. further acknowledges support from the Deutsche Forschungsgemeinschaft (German Research Foundation) under Germany’s Excellence Strategy (EXC-2123 QuantumFrontiers Grants No.~390837967) and through CRC 1227 (DQ-mat) within Project No.~A05. 
A.R. is supported by the Q-GRAV Project within the Space Research and Technology Program of the German Aerospace Center (DLR)\@. 
A.P. and E.C acknowledge support by the ``ADI 2019/2022'' project funded by the IDEX Paris-Saclay, ANR-11-IDEX-0003-02. 

Any opinions, findings, and conclusions or recommendations expressed in this article are those of the authors and do not necessarily reflect the views of the National Aeronautics and Space Administration.

\copyright 2024 All rights reserved.

\section*{Author Contributions}

J.R.W., N.P.B., and C.A.S. are CAL Principal Investigators leading the teams who conducted experiments in Sections \ref{Sect:DAIEO}, \ref{Sect:ShearWaveAI}, and \ref{Sect:PhotonRecoil}, respectively. 
The atom interferometer was proposed as a CAL add-on by the CUAS consortium including N.G., M.M., H.M., E.M.R., A.R., W.P.S. and C.Schu.\ as part of the Bigelow team. 
D.C.A. led CAL's ground testbed for the development, integration, and subsystem-level testing of the upgraded science module. 
E.R.E led operation of CAL's engineering model testbed for system-level testing of the upgraded science module with flight-like hardware. 
J.M.K supported development of the atom interferometer platform and led the characterization of instrument telemetry. 
J.R.K. led development of ISS hardware installation procedures and operations. B.S. analyzed data for the recoil measurement experiments. 
L.E.P., M.S. and S.B analyzed atom interferometry data and supported manuscript preparation. 
C.Schn.\ analyzed ISS vibration data and drafted the ISS vibrations study. 
A.R. proposed the shear-wave interferometry sequences and carried out a detailed theoretical modelling of the fringe spacing. 
H.A., P.B., M.M., G.M. and A.P., with support from E.C., E.G. and W.H., analyzed the shear-wave interferometry data. 
K.O. (CAL Project Manager) and N.E.L. led technical planning and supported testing across multiple subsystems during hardware development and science operations. 
R.J.T. proposed the CAL instrument and gave scientific guidance as Project Scientist from 2018-2020 and Cold Atom Program Scientist since 2021. 
J.R.W. was the CAL Project Scientist since 2021 and led the development of the flight atom interferometer system, with support from J.M.K., D.C.A., E.R.E., K.O. and H.M. 
The initial manuscript was drafted by J.R.W. along with C.A.S, M.M., N.G. and N.P.B.
All authors read, edited, and approved the final manuscript.

\section*{Ethics Declarations}

The authors declare no competing interests.

\section*{Data Availability Statement}
The data sets generated and analysed during the current study are available from the corresponding authors on reasonable request. All CAL data are on a schedule for public availability through the NASA Physical Science Informatics (PSI) website: \url{https://www.nasa.gov/PSI}.

\bibliographystyle{osajnl}
\bibliography{cal_ai_demo}

\end{document}